\documentstyle[12pt]{article}
\raggedright 
\renewcommand{\section}[1]{\refstepcounter{section}
\vspace{24pt}\noindent{\bf\arabic{section}.\quad #1}
\vspace*{12pt}}
\newcommand{\ulsect}[1]{\vspace{18pt}\noindent{\bf #1}
\vspace*{12pt}}

\begin{document}
\begin{flushright} CERN-TH.6813/93\\
\end{flushright}
\vspace*{10mm}
\begin{center}
{\bf Pre-equilibrium dileptons look thermal}\\[10mm]
David Seibert$^*$ and Tanguy Altherr$^{\dag}$\\[5mm] Theory
Division, CERN, CH-1211 Geneva 23, Switzerland\\[10mm]
{\bf Abstract}\\
\end{center}
\hspace*{12pt}The dilepton mass distribution from pre-equilibrium
matter in ultrarelativistic nuclear collisions is indistinguishable
from a thermally produced distribution.\\
\vfill
\begin{center}
{\em Submitted to Physics Letters B}
\end{center}
\vfill
CERN-TH.6813/93\\
February 1993\\
\vspace*{10mm}
\footnoterule
\vspace*{3pt}
$^*$On leave until October 12, 1993 from:
Physics Department, Kent State University, Kent, OH 44242 USA.
Internet: seibert@surya11.cern.ch.\\
$^{\dag}$On leave of absence from LAPP, BP 110, 74941 Annecy-le-Vieux,
France.\\
\newpage\setcounter{page}{1} \pagestyle{plain}
     \setlength{\parindent}{12pt}

Dileptons may provide one of the few windows to the early behavior
of the hot strongly-interacting matter produced in ultrarelativistic
nuclear collisions [\ref{r0}]. Because they interact only
electromagnetically, they escape from the collision volume at the
earliest stages of the collision, in contrast to strongly-interacting
particles that escape only at the end of the collision.  The dilepton
mass distribution from the equilibrating hadronic matter has recently
been calculated for the early stages of central Au+Au collisions at
$\sqrt{s}=200$ GeV/nucleon by Geiger and Kapusta [\ref{rgk}], using a
parton cascade model.  In this Letter, we show that their
distribution is indistinguishable from one produced by a thermally
equilibrated quark gluon plasma (QGP).

We calculate the dilepton mass distribution in the same manner as
Kajantie, Kapusta, McLerran and Mekjian [\ref{rkkmm}], assuming
boost-invariant longitudinal expansion and no transverse
expansion [\ref{rBj}].  In this scenario, assuming an ideal gas
of massless quarks and gluons, the temperature, $T$, can be
obtained from the proper time, $\tau$ (because entropy is
conserved):
\begin{equation}
\tau T^3 =  \mbox{\rm constant}.
\end{equation}
Thus, once the temperature is known at one proper time, it is known
at all proper times.  We fix the value of the constant from the
temperature obtained at the end of the parton cascade simulation,
$T_f=300$ MeV at $\tau_f=2$ fm/$c$ [\ref{rgk}].

The thermal dilepton mass, $M$, distribution is
\begin{equation}
\frac {d^2N} {dM^2 \, dy} ~=~ \int d\tau \, A \, \tau \,
\left. \frac {dN} {dM^2 \, d^4x} \right|_T,
\end{equation}
where $A$ is the cross-sectional area of the region of hot matter
(constant, in the absence of transverse expansion; we use $A=150$
fm$^2$).  Here
\begin{equation}
\left. \frac {dN} {dM^2 \, d^4x} \right|_T ~=~ \frac {5 \, \alpha^2}
{18 \pi^3} \, M \, T \, K_1 (M/T)
\end{equation}
is the thermal production rate from QGP [\ref{rkkmm}], where
$\alpha$ is the fine-structure constant.  We use the standard
high energy conventions that $\hbar=c=k_B=1$.

We begin our simulation at $T=T_0$, and end (as the parton
cascade calculation does) at $T_f=300$ MeV.  The total dilepton
production rate is integrated by rewriting the proper time
integrals as temperature integrals, following [\ref{rkkmm}].
\begin{equation}
\frac {d^2N} {dM^2 \, dy} ~=~ \frac {5 \, A \, \alpha^2 \,
\tau_f^2 \, T_f^6} {6 \pi^3 \, M^4} \, \left[ H(M/T_0) - H(M/T_f)
\right], \label{etdl}
\end{equation}
where
\begin{equation}
H(z) ~=~ z^2 \, (8+z^2) \, K_0(z) \, + \, 4z \, (4+z^2) \, K_1(z).
\end{equation}
The cascade results are fit extremely well by the thermal
distribution with $T_0=950$ MeV ($\tau_0=0.063$ fm/$c$), as shown
in Fig.~1.

It is obvious from the figure that the non-equilibrium production
cannot be distinguished from the equilibrium production.  The
initial proper time needed for the thermal production is very
small, much smaller than the thermalization time (0.3 fm/c)
estimated from the parton cascade calculation [\ref{rgk}].  By the
uncertainty principle, it is impossible to define the energy at
that time to an accuracy of more than about 3.1 GeV $\gg
T_0$, so the distribution of low energy quarks and gluons might
not be the same as for a thermal gas of free quarks and gluons.
However, as long as the total entropy density and the high energy
tails of the particle distributions are the same as for a free
thermal gas, the high mass dilepton production rate will be
unaffected by the non-equilibrium behavior of the low energy
components of the QGP.

At proper time $\tau_0$, the hot matter is contained in a region
of longitudinal width 0.063 fm.  The lowest mode then has energy
1.55 GeV, slightly higher than the transverse momentum cutoff,
$p_{\perp cut}=1.5$ GeV, of [\ref{rgk}].  Thus, the value of $T_0$
might be determined by the value of $p_{\perp cut}$ used in the
parton cascade model.

It is possible that this apparent thermal fit is just a coincidence;
however, it is also possible that the high energy tails of the
quark and gluon distributions are approaching thermal and chemical
equilibrium faster than the low energy quarks and gluons that
comprise the bulk of the QGP.  In either case, the result presented
here illustrates the difficulty of extracting information about the
non-equilibrium QGP from studies of high mass dileptons produced in
ultrarelativistic nuclear collisions.

\ulsect{Acknowledgement}

This material is based upon work supported by the North Atlantic Treaty
Organization under a Grant awarded in 1991.


\ulsect{References}

\begin{list}{\arabic{enumi}.\hfill}{\setlength{\topsep}{0pt}
\setlength{\partopsep}{0pt} \setlength{\itemsep}{0pt}
\setlength{\parsep}{0pt} \setlength{\leftmargin}{\labelwidth}
\setlength{\rightmargin}{0pt} \setlength{\listparindent}{0pt}
\setlength{\itemindent}{0pt} \setlength{\labelsep}{0pt}
\usecounter{enumi}}

\item E.L. Feinberg, Nuovo Cimento {\bf 34A} (1976) 391; E.V.
Shuryak, Phys.\ Lett.\ {\bf 78B} (1978) 150; G. Domokos and
J.I. Goldman, Phys.\ Rev.\ D {\bf 23} (1981) 203. \label{r0}

\item K. Geiger and J.I. Kapusta, ``Dilepton radiation from
cascading partons in ultrarelativistic nuclear collisions,''
University of Minnesota preprint (October 1992). \label{rgk}

\item K. Kajantie, J. Kapusta, L. McLerran and A. Mekjian,
Phys.\ Rev.\ D {\bf 34} (1986) 2746. \label{rkkmm}

\item J.D. Bjorken, Phys.\ Rev.\ D {\bf 27} (1983) 140.
\label{rBj}

\end{list}


\ulsect{Figure caption}

\begin{list}{\arabic{enumi}.\hfill}{\setlength{\topsep}{0pt}
\setlength{\partopsep}{0pt} \setlength{\itemsep}{0pt}
\setlength{\parsep}{0pt} \setlength{\leftmargin}{\labelwidth}
\setlength{\rightmargin}{0pt} \setlength{\listparindent}{0pt}
\setlength{\itemindent}{0pt} \setlength{\labelsep}{0pt}
\usecounter{enumi}}

\item Comparison of thermal ($T_0=950$ MeV) and non-thermal
dilepton mass distributions.

\end{list}

\vfill \eject

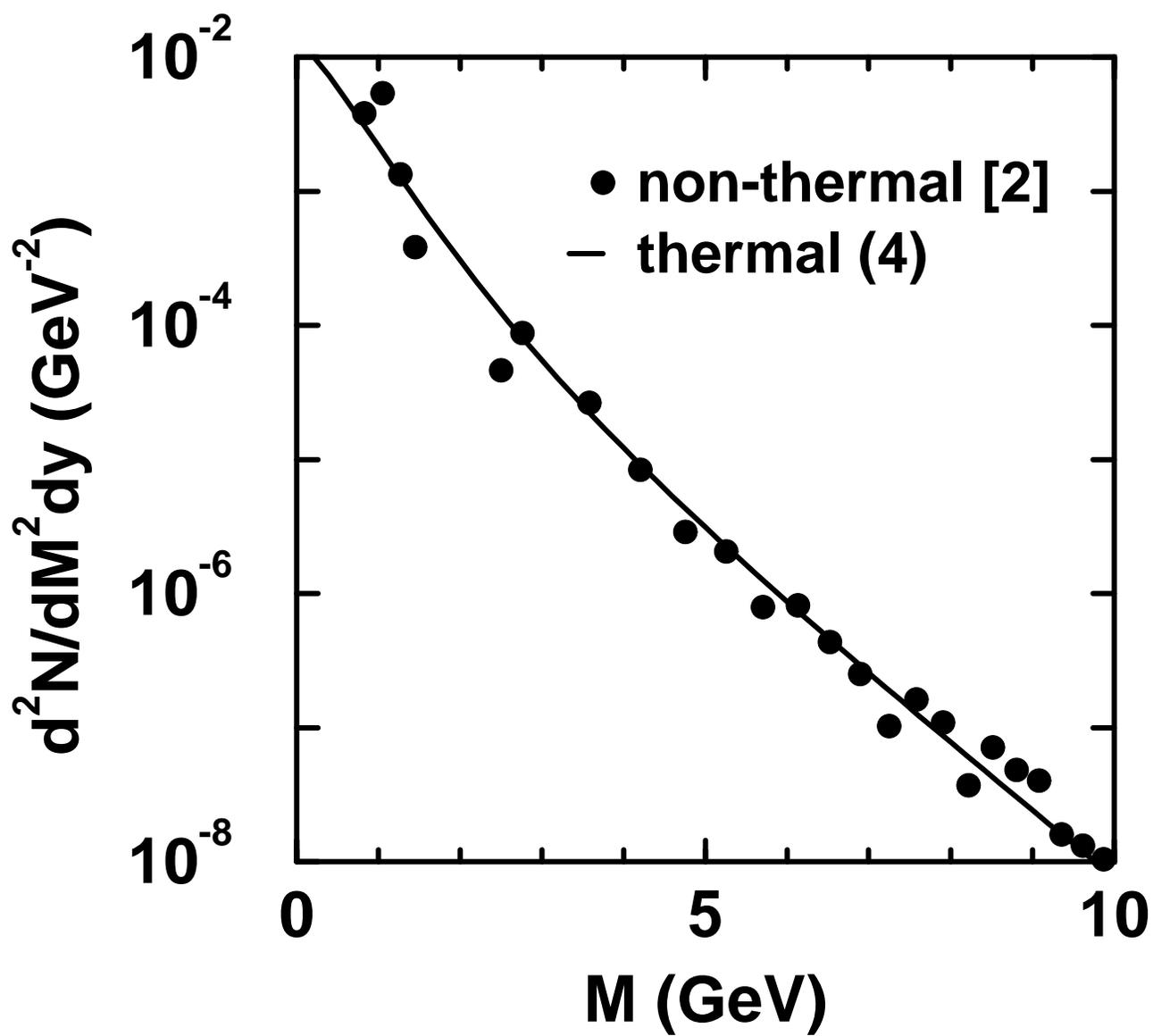
\end{document}